\documentstyle[aps,epsf]{revtex}
\newcommand{\be}{\begin{eqnarray}}
\newcommand{\ee}{\end{eqnarray}}

\draft
\begin{document}
\wideabs{
\title{Can black holes and naked singularities be detected in
accelerators?}
\author{Roberto Casadio$^{i,a}$
and
Benjamin Harms$^{ii,b}$}
\address{~}
\address{$^{i}\,$Dipartimento di Fisica, Universit\`a di
Bologna and I.N.F.N., Sezione di Bologna,\\
via Irnerio 46, 40126 Bologna, Italy}
\address{$^{ii}\,$Department of Physics and Astronomy,
The University of Alabama,\\
Box 870324, Tuscaloosa, AL 35487-0324, USA}
%
%
%
%\baselineskip 4.0ex
%\begin{titlepage}
%\pagestyle{empty}
%
%\begin{document}
\maketitle
\begin{abstract}
We study the conditions for the existence of black holes that can
be produced in colliders at TeV-scale if the space-time is higher
dimensional. On employing the microcanonical picture, we find that
their life-times strongly depend on the details of the model. If
the extra dimensions are compact (ADD model), microcanonical
deviations from thermality are in general significant near the
fundamental TeV mass and tiny black holes decay more slowly than
predicted by the canonical expression, but still fast enough to
disappear almost instantaneously. However, with one warped extra
dimension (RS model), microcanonical corrections are much larger
and tiny black holes appear to be (meta)stable. Further, if the
total charge is not zero, we argue that naked singularities do not
occur provided the electromagnetic field is strictly confined on
an infinitely thin brane. However, they might be produced in
colliders if the effective thickness of the brane is of the order
of the fundamental length scale ($\sim\,$TeV$^{-1}$).
\end{abstract}
%\end{titlepage}
}
%
%\mbox{}
%\newpage
%
%\pagestyle{plain}
%
%\raggedbottom
%\setcounter{page}{1}
%
%
\section{Introduction}
The current interest in the possibility that there exist large extra
dimensions \cite{arkani,RS} is based on the attractive features that
the hierarchy problem is by-passed by identifying the ultraviolet
cutoff with the electroweak energy scale $m_{ew}$ (without
ancillary assumptions to achieve radiative stability) and that, since
the fundamental scale of the theory is $m_{ew}$, predictions
drawn from the theory such as deviations from the $1/r^2$ law of
Newtonian gravity can be experimentally tested in the near future.
In the extra-dimensions scenarios all of the interactions, gravity
(which propagates in the whole ``bulk'' space-time) as well
gauge interactions (which are confined on the four-dimensional brane),
become unified at the electroweak scale.
This means that if the model is viable, particle accelerators such as
the CERN Large Hadron Collider (LHC), the Very LHC (VLHC)
and the Next Linear Collider (NLC) will be able to uncover the
features of quantum gravity as well as the mechanism of electroweak
symmetry breaking.
\par
The large-extra-dimension scenario also has significant implications
for processes involving strong gravitational fields, such as the
formation and decay of black holes.
Since the fundamental scale of quantum gravity is now pulled down to
order $m_{ew}$, black holes can be produced with mass of a few
TeV which behave semiclassically \cite{banks,katz}.
Since this energy scale will be available in the forthcoming
generation of colliders, they might then become black hole factories
\cite{thomas,dimopoulos}.
Black holes in $4+d$ extra dimensions have been studied in both compact
\cite{argyres,bc1,bc2} and infinitely extended \cite{chamblin} extra
dimensions (see also \cite{emparan} and references therein).
The basic feature of black hole production is that its cross section
is essentially the horizon area of the forming black hole and grows
with the center of mass energy of the colliding particles as a
power which depends on the number of extra dimensions \cite{banks}.
Although the high non-linearity of the describing equations and the
lack of a theory of quantum gravity hinder a fully satisfactory
description of this process, there are good reasons to believe in
the qualitative picture outlined above
\cite{banks,katz,thomas,dimopoulos}.
\par
Once the black hole has formed (and after a possible transient, or
``balding'' stage \cite{thomas}), Hawking radiation \cite{hawking}
is expected to set off.
The phenomenon of black hole evaporation has been described within
the context of the microcanonical ensemble in four space-time
dimensions \cite{r1,mfd} and in the context of compact extra
dimensions \cite{arkani} in Refs.~\cite{bc1,bc2}.
Our starting point is the idea that black holes are (excitations of)
extended objects ($p$-branes), a gas of which satisfies the bootstrap
condition.
This yields a picture in which a black hole and the emitted particles
are of the same nature and an improved law of black hole decay which is
consistent with unitarity (energy conservation).
\par
The use of the microcanonical picture will lead us to the conclusion
that the evaporation process in the presence of extra dimensions
strongly depends on the details of the model.
In particular, if the extra dimensions are compact (ADD scenario
of Ref.~\cite{arkani}) the luminosity of tiny black holes is in poor
qualitative agreement with that predicted by the canonical picture
since the occupation number density departs from thermality for
masses slightly above the TeV-scale.
On applying the formalism of Ref.~\cite{bc2} to the cases of
interest, we shall then argue that a black hole produced in a
collider would be relatively longer-lived with respect
to estimates in the existing literature \cite{thomas,dimopoulos}.
However, the typical life-time is short enough that black holes
can be considered to decay (at least down to the fundamental mass
scale) instantaneously.
On the other hand, if there is one warped extra dimension
(RS scenario of Ref.~\cite{RS}), the microcanonical luminosity
differs significantly from the canonical expression, and
the evaporation process might be frozen below the scale at which
corrections to Newton's law become effective.
\par
It is also important to note that such tiny singularities in four
dimensions, besides being beyond the realm of classical general
relativity, would be black holes only provided their electric
charge is zero, otherwise they are naked singularities. In the
following Section we shall consider such cases. We know of no
conclusive argument which completely rules out their existence.
\par
We shall use units with $c=1$, $\hbar=m_p\,l_p$
($l_p$ is the four-dimensional Planck length) and
$G_N=l_p/m_p$ denotes the four-dimensional Newton constant.
\section{Naked singularities}
\label{nak}
The four-dimensional argument about naked singularities mentioned
at the end of the Introduction easily generalizes to higher
dimensions.
In fact, one observes that charged (spherically symmetric) black
holes must satisfy the inequality \cite{myers}
\be
Q^2\,{m_p^2\over M^2}<{(2+d)\,(1+d)\over 2}
\ ,
\label{naked}
\ee
where $Q$ is the charge in dimensionless units.
The condition in Eq.~(\ref{naked}) is obviously violated in the
ADD scenario, where the effect of the brane on the space-time
geometry is essentially neglected, since an object with mass
of order a few TeV and charge equal to (fractions of) the
electron charge has $Q\,(m_p/M)\sim 10^8$.
\par
A possible way to circumvent the bound (\ref{naked}) is by
requiring that the electromagnetic field of a point-like charge be
confined to the brane, thereby, spoiling the full
$(3+d)$-dimensional spherical symmetry \cite{landsberg}.
The system would thus appear spherically symmetric only from the
four-dimensional point of view.
The only known metric on the brane which might represent such a
case was found in Ref.~\cite{maartens} in the context of the RS
scenario~\footnote{Its extension into the bulk is still under study
(see, e.g., Ref~\cite{shiromizu} for a numerical analysis).}.
Such a solution has the Reissner-Nordstr\"om form
\be
-g_{tt}={1\over g_{rr}}=
1-2\,{M\,l_p\over m_p\,r} +Q^2\,{l_p^2\over r^2}
-q\,{m_p^2\,l_p^2\over m_{(5)}^2\,r^2}
\ ,
\ee
and the (outer) horizon radius is given by
\be
R_H=l_p\,{M\over m_p}\,
\left[1+\sqrt{1-Q^2\,{m_p^2\over M^2}
+{q\,m_p^4\over M^2\,m_{(5)}^2}}\right]
\ ,
\label{RH_RS}
\ee
where $m_{(5)}\sim m_{ew}$ is the fundamental mass scale and
$q$ represents a (dimensionless) tidal charge.
The latter can be estimated on dimensional grounds as
\cite{maartens,bc2}
\be
q\sim\left({m_p\over m_{ew}}\right)^\alpha\,{M\over m_{ew}}
\ ,
\ee
and for $\alpha>-4$ the tidal term $\sim 1/r^2$ dominates over the
four-dimensional potential $\sim 1/r$ (as one would expect for
tiny black holes).
The condition (\ref{naked}) is therefore replaced by
\be
Q^2\,{m_p^2\over M^2}<
1+\left({m_p\over m_{ew}}\right)^{3+\alpha}\,{m_p\over M}
\ ,
\label{naked1}
\ee
which can be fairly large for $\alpha>-4$ and $M\sim m_{ew}$,
in contrast to the right hand side of Eq.~(\ref{naked}).
\par
Which of the two conditions (\ref{naked}) and
(\ref{naked1}) is relevant remains an open question,
since one might in fact argue that the brane cannot be
infinitely thin (see, e.g., \cite{cgv} and Refs. therein).
Gauge fields would then extend along the extra dimension(s),
roughly to a width of the order of $m_{ew}^{-1}$, and
this likely yields a bound somewhere in between the expressions
given in Eq.~(\ref{naked}) and Eq.~(\ref{naked1}) for a
singularity with $R_H\sim m_{ew}^{-1}$.
There is thus no compelling reason to discard the possibility
that the collision of charged particles produces a naked
singularity, an event which would probably be indistinguishable
from ordinary particle production, with the naked singularity
(possibly) behaving as an intermediate, highly unstable state.
The phenomenology of naked singularities is probably rather
different from that of black holes, as they are generally expected
to explode in a very sudden event instead of evaporating via the
Hawking process (at least in an early stage; see, e.g.,
\cite{harada} and Refs.~therein).
\par
We should however add that the present literature does not
reliably cover the case of such tiny naked singularities
and their actual phenomenology is an open question.
A naked singularity is basically a failure in the
causality structure of space-time mathematically admitted
by the field equations of general relativity.
Most studies have thus focused on their realization as the
(classical) end-point of the gravitational collapse of compact
objects (such as dust clouds) and on their stability by
employing quantum field theory on the resulting background.
However, one might need more than semiclassical tools
to investigate both the formation by collison of particles
and the subsequent time evolution \cite{harada}.
In particular, to our knowledge, no estimate of the life-time
of a naked singularity of the sort of interest here is yet
available.
\par
To summarize, the following two cases might occur in a proton-proton
collider such as the LHC,
\be
p^++p^+\ \rightarrow\
\left\{\begin{array}{l}
{\rm B.\,H.}+X^{++}
\\
\\
{\rm N.\,S.\ \ or\ \ B.\,H.}+Y^{0,+}
\ ,
\end{array}
\right.
\ee
where $X^{++}$ denotes a set of particles whose total charge
is twice the proton charge and $Y^{0,+}$ a set of particles with
vanishing total charge or with one net positive charge.
\section{Black holes}
In a four-dimensional space-time, a black hole might emerge from
the collision of two particles only if its center of mass energy
exceeds the Planck mass $m_p$. In fact, $m_p$ is the minimum mass
for which the Compton wavelength $l_M=l_p\,(m_p/M)$ of a
point-like particle of mass $M$ equals its gravitational radius
$R_H=2\,G_N\,M$. For energies below $m_p$ the very (classical)
concept of a black hole would lose its meaning. However, since the
fundamental mass scale is shifted down to $m_{ew}$ in the models
under consideration, black holes with $M\ll m_p$ can now exist as
classical objects provided
\be
l_p\,{m_p\over M}\ll R_H\ll L
\ ,
\label{class}
\ee
where $L$ is the scale at which corrections to
the Newtonian potential become effective. The left hand inequality
ensures that the black hole behaves semiclassically, and one does
not need a full-fledged theory of quantum gravity, while the right
hand inequality guarantees that the black hole is small enough
that its gravitational field can depart from the Newtonian
behavior without contradicting present experiments. In this
Section we check that black holes with $m_{ew}<M<10\,m_{ew}$ are
allowed and then study their evaporation process. We shall have
nothing new to report about the cross section for their
production.
\par
The luminosity of a black hole in $D$ space-time dimensions is
given by
\be
{\mathcal L}_{(D)}(M)=
{\cal A}_{(D)}\,\int_0^{\infty}\sum_{s=1}^S\,
n_{(D)}\,(\omega)\,\Gamma_{(D)}^{(s)}(\omega)\,\omega^{D-1}\,
d\omega
\label{dMdt}
\ee
where ${\mathcal A}_{(D)}$ is the horizon
area in $D$ space-time dimensions, $\Gamma_{(D)}^{(s)}$ the
corresponding grey-body factor and $S$ the number of species of
particles that can be emitted.
For the sake of simplicity, we shall approximate
$\sum_s\,\Gamma^{(s)}_{(D)}$ as a constant (see
Section~II.C in \cite{bc2} and below).
The distribution $n_{D}$ is the microcanonical number density
\cite{r1,mfd}
\be
n_{(D)}(\omega)=C\sum_{l=1}^{[[M/\omega]]}
\exp\left[S_{(D)}^E(M-l\,\omega)-S_{(D)}^E(M)\right]
\label{n}
\ee
where $[[X]]$ denotes the integer part of $X$ and $C=C(\omega)$
encodes deviations from the area law \cite{r1} (in the following
we shall also assume $C$ is a constant in the range of interesting
values of $M$).
The basic quantity in Eq.~(\ref{n}) is the Euclidean black hole
action, which usually takes the form
\be
S_{(D)}^E\sim{{\mathcal A}_{(D)}\over l_{(D)}^2} =\left({M\over
m_{eff}}\right)^\beta
\ ,
\label{action}
\ee
where $m_{eff}$ and $\beta$ are model-dependent quantities and
$l_{(D)}$ ($m_{(D)}$) is the fundamental length (mass) in $D$
space-time dimensions related to the fundamental Newton constant
by
\be
G_{(D)}={l_{(D)}^{D-3}\over m_{(D)}}
\ .
\ee
We recall that for $\beta=\beta_d\equiv(2+d)/(1+d)$ and in the limit
$M/m_{eff}\to\infty$, $n_{(4+d)}(\omega)$ mimics the canonical
ensemble (Planckian) number density in $4+d$ space-time dimensions
and the luminosity becomes
\be
{\mathcal L}_{(4+d)}^H\sim{\mathcal A}_{(4+d)}\,
(T^H_{(4+d)})^{4+d} \sim{1\over R_H^2}
\ ,
\label{L_H}
\ee
where $T^H_{(4+d)}$ is the Hawking temperature in $4+d$
dimensions.
\par
On using Eqs.~(\ref{n}) and (\ref{action}) one can show that
the luminosity is in general given by
\be
{\mathcal L}_{(4+d)}=K\,
m^\beta\,e^{-m^\beta}\,\int_0^m
e^{x^\beta}\,\left(m-x\right)^{3+d}\,dx
\ ,
\label{dMdteff}
\ee
where $m\equiv M/m_{eff}$ and $K$ is a coefficient which
contains all the dimensionful parameters but does not
depend on $M$.
The above integral can be performed exactly for the
models under consideration.
\par
We shall now analyze the ADD and RS scenarios separately.
\subsection{ADD scenario}
If the space-time is higher dimensional and the extra dimensions
are compact and of size $L$, the relation between the mass of a
spherically symmetric black hole and its horizon radius is changed
to \cite{myers}
\be
R_H\simeq l_{(4+d)}\,
\left({2\,M\over m_{(4+d)}}\right)^{1\over 1+d}
\ ,
\label{R_H<}
\ee
where
\be
G_{(4+d)}\simeq L^d\,G_N
\ ,
\ee
is the fundamental gravitational constant in $4+d$ dimensions.
Eq.~(\ref{R_H<}) holds true for black holes of size $R_H\ll L$,
or, equivalently, of mass
\be
M\ll M_c\equiv m_p\,{L\over l_p}
\ .
\ee
Since $L$ is related to $d$ and the fundamental mass scale
$m_{(4+d)}\sim m_{ew}\sim 1\,$TeV by \cite{arkani}
\be
{L\over l_p}\sim
\left[{m_{ew}\over m_{(4+d)}}\right]^{1+{2\over d}}\,
10^{{31\over d}+16}
\equiv
\gamma^{1+{2\over d}}\, 10^{{31\over d}+16}
\ ,
\label{tev}
\ee
Eq.~(\ref{class}) translates into
\be
10^{-{31+16\,d\over 2+d}}\,\gamma\,m_p
\sim 10^{-16}\,\gamma\,m_p
\ll M\ll M_c
\ ,
\ee
\begin{figure}
\centering
\raisebox{4cm}{${\mathcal L}_{(10)}$}
\epsfxsize=2.9in
\epsfbox{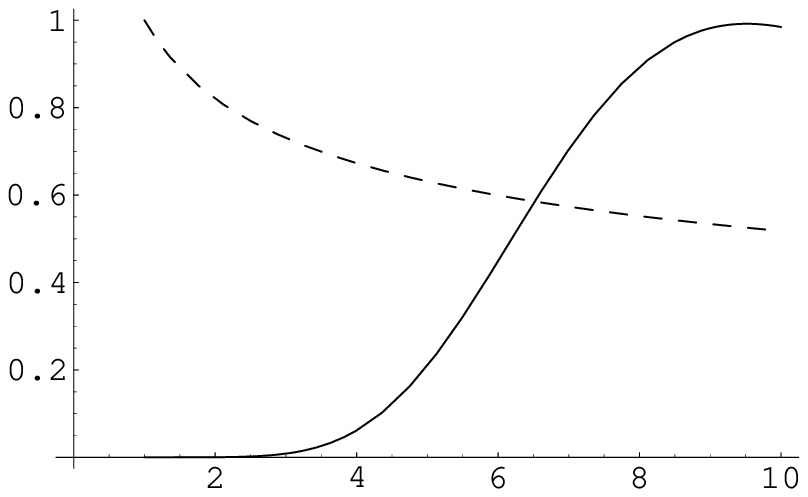}
\hspace{-0.2in}
\raisebox{0.5cm}
{\hspace{6cm} ${m\over m_{ew}}$}
\caption{Microcanonical luminosity (solid line) for a small
black hole with $d=6$ extra dimensions compared to the
corresponding canonical luminosity (dashed line).
Vertical units are chosen such that
${\mathcal L}^H_{(10)}(m_{ew})=1$.}
\label{L_add}
\end{figure}
\par\noindent
where we also used the fact that $d=1$ is ruled out by present
measurement of $G_N$ \cite{arkani} and relatively high values
of $d$ (i.e., $d\sim 6$) seem to be favored
(see, e.g., Refs.~\cite{bounds}).
For $\gamma\sim 1$ (i.e., $m_{(4+d)}$ of order
$m_{ew}\sim 1\,$TeV),
the left hand side above is of order $m_{ew}$ as well.
\par
The Euclidean action is of the form in Eq.~(\ref{action}) with
$m_{eff}=m_{(4+d)}\sim m_{ew}$ and $\beta_d=(d+2)/(d+1)$.
The occupation number density for the Hawking particles in the
microcanonical ensemble is thus given by
\be
n_{(4+d)}(\omega)\sim\sum_{l=1}^{[[M/\omega]]}\,
e^{\left({M-l\,\omega\over m_{ew}}\right)^{d+2\over d+1}
-\left({M\over m_{ew}}\right)^{d+2\over d+1}}
\ .
\label{n_add}
\ee
We then notice that the above expression differs from the
four-dimensional one for which $m_{eff}=m_p\gg m_{ew}$ and
$\beta=2>\beta_d$.
In four dimensions one knows that microcanonical corrections
to the luminosity become effective only for $M\sim m_p$,
therefore, for black holes with $M\gg m_{ew}$ the luminosity
(\ref{dMdt}) should reduce to the canonical result given
in Eq.~(\ref{L_H})
\cite{emparan,bc1,bc2}~\footnote{This was also shown to be a
good approximation of the luminosity as seen by an observer
on the brane, since most of the emission occurs into particles
confined on the brane \cite{emparan,bc1}.}.
In order to eliminate the factor $K$ from Eq.~(\ref{dMdteff}),
one can therefore equate the microcanonical luminosity to the
canonical expression at a given reference mass
$M_0\sim M_c\gg m_{ew}$ and then normalize the microcanonical
luminosity according to
\be
{\mathcal L}_{(4+d)}(M)&\simeq&
{{\mathcal L}_{(4+d)}^H(M_0)
\over
{\mathcal L}_{(4+d)}(M_0)}\,
{\mathcal L}_{(4+d)}(M)
\ .
\label{L_norm}
\ee
The black hole luminosity thus obtained differs significantly
from the canonical one for $M\sim m_{ew}$, as can be clearly
seen from the plot for $d=6$ in Fig.~\ref{L_add}.
For smaller values of $d$ the picture remains qualitatively the
same, except that the peak in the microcanonical luminosity
shifts to lower values of $M$ and this affects the
ratio
\be
{\mathcal R}_{(4+d)}(m_{ew})\equiv
{{\mathcal L}_{(4+d)}(m_{ew})\over
{\mathcal L}_{(4+d)}^H(m_{ew})}
\ .
\label{rat}
\ee
Although the integral in Eq.~(\ref{dMdteff}) can be performed
exactly, its expression is very complicated and we omit it.
Instead, in Table~\ref{t1} we show the relevant quantities for
$d=2,\ldots,6$ (upper bounds for the grey-body factors are
estimated as in Ref.~\cite{bc2} for $s$-wave modes only, since
one expects significantly smaller values for non-zero angular
momentum modes \cite{page}).
\par
In all cases, the microcanonical luminosity becomes smaller
for $M\sim m_{ew}$ than it would be according to the
canonical luminosity, which makes the life-time of the
black hole somewhat longer than in the canonical
picture.
In particular, for $d=6$ one finds
\be
\left.{dM\over dt}\right|_{M\sim m_{ew}}
\simeq -10^{-10}\,{\mathcal L}_{(10)}^H
\sim -10^{17}\,{{\rm TeV}\over {\rm s}}
\ .
\ee
In the range $6\,$TeV$<M<10\,$TeV the luminosity
is actually larger than the canonical expression
(see Fig.~\ref{L_add}).
A black hole would therefore evaporate very quickly
\cite{dimopoulos} down to $\sim 6\,m_{ew}$.
Then, its life-time is dominated by the time it would
take to emit the remaining $\Delta M\sim 5\,$TeV,
before it reaches $1\,m_{ew}$, which is approximately
\be
T\sim \left({dM\over dt}\right)^{-1}\,\Delta M
\sim 10^{-17}\,{\rm s}
\ .
\label{T}
\ee
\par
The above relatively long time does take into account
the dependence of the grey-body factor $\Gamma_{(4+d)}^{(s)}$
on $d$ but not the actual number $S$ of particle species into
which the black hole can decay.
The latter would increase the luminosity by a factor
$S\sim 10\to 100$ \cite{dimopoulos}, but this is already taken
care of by the normalizing procedure defined by Eq.~(\ref{L_norm}).
One might actually guess that the number $S$ in the microcanonical
picture is smaller than in the canonical framework, since
the ``effective'' canonical temperature of tiny black holes is much
smaller than the corresponding $T^H_{(4+d)}$.
Hence we conclude that the value given in Eq.~(\ref{T}) is a
fairly good estimate of the order of magnitude of the true
life-time.
It is of course quite small with respect to the sensitivity
of present detectors, which is on the order of hundreds of
picoseconds.
\par
In the above analysis we have only considered masses for a
black hole larger than the fundamental $m_{ew}\sim 1\,$TeV
scale.
One can assume that, once the fundamental scale has been reached,
a black hole continues to evaporate.
However, it is also possible that the radiation simply switches
off at that point (as the microcanonical luminosity suggests)
and the small black hole escapes as a stable remnant.
If heavy ($\sim 10\,$TeV) black holes are produced, they will
be moving slowly and will quite likely decay (at least down to
about $1\,$TeV) in the detector producing a ``sudden burst''
of light particles (electrons, positrons, neutrinos and
$\gamma$-rays) at the collision point.
However the cross section for the production of such heavy
black holes is very small \cite{dimopoulos}.
If a neutral black hole is produced with a mass
$\sim m_{ew}$ and is stable, its detection will
depend upon the ability of the detectors to measure the missing
transverse momentum or the missing mass accurately enough
to prove the existence of a massive neutral particle.
If instead it is charged and stable, it should not be
difficult to track its path.
Limits to the existence of stable remnants should also come,
e.g., from estimates of the allowed density of primordial
black holes \cite{carr}.
\subsection{RS scenario}
In order to study this case, we shall again make use of the
solution given in Ref.~\cite{maartens} (although new metrics were
given in Ref.~\cite{cfm}).
From Eq.~(\ref{RH_RS}) with $\tilde Q=0$ and $\alpha>-4$ one
obtains
\be
R_H\simeq
l_p\,\left({m_p\over m_{(5)}}\right)^{1+{\alpha\over 2}}\,
\sqrt{M\over m_{(5)}}
\ ,
\ee
since the tidal term $q$ dominates for both $M$ and
$m_{(5)}\ll m_p$, and one must still have Eq.~(\ref{class}).
With one warped extra dimension \cite{RS}, the
length $L$ is just bounded by requiring that Newton's law not be
violated in the tested regions, since corrections to the $1/r$
behavior are of order $(L/r)^2$.
This roughly constrains $l_p<L<10^{-3}\,$cm.
Hence the allowed masses are, according to
Eq.~(\ref{class}),
\be
\left({m_{(5)}\over m_{p}}\right)^{{\alpha\over 3}}
\ll {M\over m_{(5)}}
\ll\left({L\over l_p}\right)^{2}\,
\left({m_{(5)}\over m_p}\right)^{2+\alpha}
\ .
\ee
In particular one notices that black holes with
$M\sim m_{(5)}\sim m_{ew}$ could exist only if
the following two conditions are simultaneously satisfied
\be
\alpha\ge 0
\ \ \ \ {\rm and}
\ \ \ \ {L\over l_p}\gg
\left({m_p\over m_{ew}}\right)^{3+\alpha\over 3}
\ .
\label{exist}
\ee
\par
The luminosity is now given by the four-dimensional expression
(\ref{dMdteff}) with $D=4$, $\beta=1$ and
\be
m_{eff}=\left({m_{ew}\over m_p}\right)^{2+\alpha}\,m_{ew}
\ .
\ee
The result is simple enough to display, namely
\be
{\mathcal L}_{(4)}&=&K\,m\,e^{-m}\,\left(e^m-{1\over 6}\,m^3
-{1\over 2}\,m^2-m-1\right)
\nonumber \\
&\simeq&K\,m
\ ,
\ee
where the last expression follows from $m=M/m_{eff}\ll 1$ since
$m_{eff}\ll m_{ew}$ for $\alpha\ge 0$.
We again eliminate $K$ by normalizing the luminosity to the
(four-dimensional) canonical expression
${\mathcal L}_{(4)}^H(M_0)$, where now $M_0\sim M_c=m_p\,(L/l_p)$
is the mass above which corrections to Newton's law are negligible.
For the limiting case $\alpha=0$, on taking into account the second
condition in Eq.~(\ref{exist}) one obtains
\be
{\mathcal L}_{(4)}<
10^{-9}\,{M\over m_{ew}}\,{{\rm TeV}\over{\rm s}}
\ ,
\ee
which yields an exponential decay with typical life-time
$T>10^9\,$s.
\par
The above result is certainly striking, since it means
that microscopic black holes are (meta)stable objects and would be
detected just as missing energy (if neutral) or stable heavy
particles (if charged).
Hence, either they escape from the detector and carry away
a large amount of energy or in rare instances they give rise to
an isotropic (almost steady) vanishingly faint flux of particles
(a ``star'') inside the detector.
Black holes with life-times this long would have had an effect on
the evolution of the early universe.
The allowed density of primordial black holes \cite{carr} might
thus be able to provide some evidence as to the validity of the
RS scenario.
\section{Conclusions}
We have analyzed the conditions for the existence of naked
singularities and black holes with masses that can be reached in
accelerators such as the LHC. We have shown that the typical
life-times of tiny black holes depend strongly on the model
employed, since they decay almost instantaneously in ADD and are
(quasi)stable in RS. As we argued in Section~\ref{nak} for the
existence of naked singularities, it is likely that the results in
RS apply only to a brane-world of zero thickness. For a brane of
width of order TeV$^{-1}$, the correct life-times would probably be in
between those predicted in ADD (where the brane is
totally neglected) and those in RS. We think this shows that the
phenomenology of such objects is not completely settled,
mainly due to the persistent lack of a sensible solution
representing a black hole in a space-time with extra dimensions in
which our brane-world would be embedded.
\begin{table}
\centering
\begin{tabular}{|c|c|c|c|c|}
\hline
$d$
& ${\mathcal L}_{(4+d)}^H$
& $\Gamma_{(4+d)}$
& ${\mathcal R}_{(4+d)}$
& ${\mathcal L}_{(4+d)}$
\\
\hline
$2$
& $10^{28}$ & $10^{-1}$ & $10^{-2}$ & $10^{25}$
\\
\hline
$3$
& $10^{28}$ & $10^{-1}$ & $10^{-4}$ & $10^{23}$
\\
\hline
$4$
& $10^{28}$ & $10^{-2}$ & $10^{-5}$ & $10^{21}$
\\
\hline
$5$
& $10^{28}$ & $10^{-2}$ & $10^{-6}$ & $10^{20}$
\\
\hline
$6$
& $10^{27}$ & $10^{-3}$ & $10^{-7}$ & $10^{17}$
\\
\hline
\end{tabular}
\caption{
Relevant quantities for the ADD scenario.
${\mathcal L}_{(4+d)}^H$ is the canonical luminosity in
TeV$/$s;
$\Gamma_{(4+d)}$ is an upper bound for the grey-body factor;
${\mathcal R}_{(4+d)}$ is the ratio defined in Eq.~(\ref{rat})
and ${\mathcal L}_{(4+d)}$ the microcanonical luminosity
in TeV$/$s.
All quantities are evaluated for $M\sim m_{ew}\sim 1\,$TeV.}
\label{t1}
\end{table}
\acknowledgments
We wish to acknowledge useful discussions with J.~Busenitz,
L.~Clavelli and G.~Landsberg.
This work was supported in part by the U.S. Department of Energy
under Grant no.~DE-FG02-96ER40967.
\end{document}